\documentclass[11pt, a4paper]{article}
\pdfoutput=1
\usepackage{jheppub}
\usepackage{graphicx}
\usepackage{epsfig,amsmath,amsthm}
\usepackage{epstopdf}
\newcommand{\be}{\begin{equation}}
\newcommand{\ee}{\end{equation}}
\newcommand{\bea}{\begin{eqnarray}}
\newcommand{\eea}{\end{eqnarray}}
\def\bse{\begin{subequations}}
\def\ese{\end{subequations}}

\def\IZ{\relax\ifmmode\hbox{Z\kern-.4em Z}\else{Z\kern-.4em Z}\fi}

\newcommand{\non}{\nonumber \\}

\def\half{\frac{1}{2}} 
\def\del{{\partial}}

\DeclareMathOperator\sgn{sgn}

\def\bGm{{\bar \Gamma}}
 
\def\cE{{\cal E}}
\def\tlf{\tilde{l}_F}
 
  \def\eps{\epsilon}
\def\sig{\sigma} \def\bsig{{\bar \sigma}}

\def\presub{\vspace{.5cm} \noindent}

\def\bi{\begin{itemize}} \def\ei{\end{itemize}}

\def\({\left(} \def\){\right)}
\def\[{\left[} \def\]{\right]}
\def\<{\left<} \def\>{\right>}

\usepackage{tikz}
\newcommand{\dittotikz}{%
    \tikz{
        \draw [line width=0.12ex] (-0.2ex,0) -- +(0,0.8ex)
            (0.2ex,0) -- +(0,0.8ex);
        \draw [line width=0.08ex] (-0.6ex,0.4ex) -- +(-1.5em,0)
            (0.6ex,0.4ex) -- +(1.5em,0);
    }%
}

\title{Flux-based statistical prediction of three-body outcomes}

\author{Barak Kol  \\
{\it Racah Institute of Physics, Hebrew University, Jerusalem 91904, Israel} \\
{\tt barak.kol@mail.huji.ac.il}
}

\abstract{The gravitational three-body problem is a rich open problem, dating back to Newton. It serves as a prototypical example of a chaotic system and has numerous applications in astrophysics. Generically, the motion is non-integrable and susceptible to disintegration, and for negative total energy the decay outcome is a free body flying apart from a binary. Since Poincar\'e, the problem is known to be chaotic and is believed to lack a general deterministic solution. Instead, decades ago a statistical solution was marked as a goal. Yet, despite considerable progress, all extant approaches display two flaws. First, probability was equated with phase space volume, thereby ignoring the fact that significant regions of phase space describe regular motion, including post-decay motion. Secondly and relatedly, an adjustable parameter, the strong interaction region, which is a sort of cutoff, was a central ingredient of the theory. 

This paper introduces remedies and presents for the first time a statistical prediction of decay rates, in addition to outcomes. Based on an analogy with a particle moving within a leaky container, the statistical distribution is presented in an exactly factorized form. One factor is the flux of phase-space volume, rather than the volume itself, and it is given in a cutoff-independent closed-form. The other factors are the chaotic absorptivity and the regularized phase space volume. The situation is analogous to Kirchhoff's law of thermal radiation, also known as greybody radiation. In addition, an equation system for the time evolution of the statistical distribution is introduced; it describes the decay rate statistics while accounting for sub-escape excursions. Early numerical tests indicate a leap in accuracy. The author believes that the results of this paper are an essential ingredient of the envisioned statistical theory, that they may very well affect our understanding of related chaotic systems, and that they would have astrophysical applications.}

\begin{document}

\maketitle

\section{Introduction}

The three body problem is concerned with the study of the motion of three point masses influenced by their mutual gravitational attractions.

This problem has a long and venerable history. It was appreciated already by Newton and discussed in his 1687 Principia \cite{Principia} 333 years ago. Its name became common in the 1740s in connection with a rivalry between d'Alembert and Clairaut who studied it. Euler \cite{Euler_1767} and Lagrange \cite{Lagrange_1772} obtained special periodic solutions, in 1767 and 1772 respectively, and their restriction became known as Lagrangian points. 

Important work was done on the problem's perturbative limit, where masses or orbit sizes are hierarchical, but this paper focuses on the generic problem.

By 1887 the problem had become so celebrated that a prize was offered for its solution by Oscar II, king of Sweden, who was advised by the mathematician Mittag-Leffler.  Poincar\'e 
 accepted this challenge and after first presenting a faulty submission finally came to realize that the problem exhibits a sensitivity to initial conditions, in the sense that a small change in initial conditions leads to a quickly increasing difference between the two solutions \cite{Poincare_1892},\cite{Barrow-Green_1996}.  This makes a general analytic solution impossible and breaks the mechanical paradigm that knowledge of a system's forces and initial conditions allows us to predict its motion at arbitrary times in the future. Since the initial conditions are always known only up to some accuracy, after some time the prediction error increases so much that the prediction loses its value. In modern terms this was the first example of a non-integrable, chaotic system, see e. g. \cite{Hand_Finch_1998}. Moreover, this realization won Poincar\'e the prize. At the beginning of the 20th century, this was the status of the three-body problem.
 
While it is impossible to predict the general motion of the three body system into the far future, it still makes sense to ask what would be the \emph{likely} result of any given initial conditions. This suggests to seek a statistical solution. This is not the usual Statistical Physics approach to thermodynamics which holds for a large collection of particles, see e.g. \cite{LL_stat_phys}, but rather a statistical mechanics approach where one considers an ensemble of initial conditions for a system with few degrees of freedom, as advocated by Gibbs \cite{Gibbs1902}. Moreover, it is natural to define probability to be proportional to phase space volume, namely, an ergodic approximation. In general, the more chaotic a system is, the shorter is the validity of a simulated trajectory, while at the same time the statistical analysis becomes more accurate. In this way, the non-integrability evolves from a liability into an advantage.  

The development of computers and computational physics allowed us to integrate the three-body equations of motion numerically  \cite{Agekyan_Anosova_1967,Standish_1972,Saslaw_Valtonen_Aarseth_1974} and to do so for large numbers of sets of initial conditions. Motion with negative total energy was found to generally result in an escape of one of the masses. This need not be a surprise since such an outcome is allowed by conservation laws, and the chaotic nature suggests that it would indeed be realized. 

The possible end-states, or outcomes, are parameterized by several variables. In 1976 Monaghan \cite{Monaghan1_1976} suggested applying the statistical approach to the three-body problem, and within the ergodic approximation defined the outcome probability distribution in terms of appropriate phase space volumes. This approach was developed in \cite{Monaghan2_1976,Monaghan3_1978} and elsewhere, see the wonderful books \cite{Valtonen_book_2006,Valtonen_etal_book_2016}, the review \cite{Musielak_Quarles_rev_2014} and references therein. Recently, Stone and Leigh \cite{Stone_Leigh_2019} presented an outcome distribution in closed-form by improving the evaluation of phase space volumes using canonical transformations to Delaunay elements.  

Apart from being a leading example for chaotic dynamics, the system is linked to numerous astrophysical phenomena, see e.g. the last paragraph of \cite{Stone_Leigh_2019} and references therein. In particular, it is involved in a mechanism suggested to produce some of the tight binaries detected by gravitational wave observatories.

Despite considerable progress, all extant approaches are afflicted by certain flaws and incompleteness. First, the identification of probability with phase space volume ignores the fact that a considerable part of phase space describes regular motion. In particular, the generic motion is not bounded, but rather leads to decay,  and once the system separates into a single and a binary flying away from each other, the motion is regular. In fact, it even becomes free once they are sufficiently far away. Secondly, the extant approaches introduce a spurious parameter, the strong interaction region, which prevents infinite phase volumes from appearing and limits the contamination of phase space volumes by regions of regular motion.  Finally, extant approaches attempt to model the outcome distribution, but not the decay time, and in this sense are incomplete.

This paper's research objective is to formulate a comprehensive statistical prediction, free of the above described flaws and incompleteness. Such a prediction would address not only the outcome distribution, but also the decay time and its statistics, while accounting for regions of regularity and the unbounded nature of the motion without introducing a spurious parameter.

The first main idea of this paper is that the chaotic decay of three-body motion is similar to a particle moving inside a container with a perfectly reflecting wall that has a small hole in it.  Here also, one expects that general trajectories will be chaotic and will end in escape from the container. In fact, the decay rate is given by the flux of phase space volume (or phase-volume) throughout the hole, divided by the total phase-volume within the container. This clarifies that the phase-volume element is not the correct measure of probability outside the container. In order to implement the analogy, we need to define the phase-volume flux and the total phase-volume for the three-body problem, which, unlike the container, is an unbounded system. 

The second main idea is to account for regular trajectories through a factorization of the flux out of the chaotic region into the total outgoing flux and the chaotic absorptivity. 
 It is analogous to Kirchhoff's law of thermal radiation ($\sim$1860) which equates emissivity with absorptivity (the greybody factor is another, equivalent, term). In fact, we believe ergodic factorization implies it.

Thirdly, we account for sub-escape excursions through a statistical evolution model. In it, the distribution is divided into several compartments, represented by a piping diagram where the main compartments are the chaotic region, sub-escape excursions and escapes. The time evolution of the distribution is shown to satisfy an equation system, where the differential decay rate is central.

Section \ref{sec:gen_pred} sets up the problem and presents the main results which are the interconnected expressions for the total outgoing flux, the differential decay rate and the statistical evolution. Section \ref{sec:derive} presents motivations and derivations.  
On the way, we present a new method for integration of micro-canonical phase-volumes over phase space momenta. Section \ref{sec:spec_pred} presents more predictions in preparation for a comparison with numerical integration. We conclude in section \ref{sec:discussion} with a summary and discussion. An appendix includes the marginal distributions within some crude, yet useful, approximation.

\newpage
\section{Flux-based statistical prediction}
\label{sec:gen_pred}

\subsection{Setup and outcome parameters}

The gravitational three-body problem consists of the motion of three point masses $m_1,\, m_2,\, m_3$ influenced by their mutual gravitational interaction. The system can be defined through the Hamiltonian 
\bea
	H\( \{\vec{r}_c,\, \vec{p}_c \}_{c=1}^3 \) &:=& T+V \label{def:H} \\
	T &:=&  \sum_{c=1}^3 \frac{\vec{p}_c^{~2}}{2 m_c} \non
	V &:=& - \frac{G\, m_1\, m_2}{r_{12}} -  \frac{G\, m_1\, m_3}{r_{13}} - \frac{G\, m_2\, m_3}{r_{23}} \nonumber
 \eea 
 where $\vec{r}_c, ~c=1,2,3$ are the bodies' position vectors, $\vec{p}_c$ are their momenta, $G$ is Newton's gravitational constant, and $r_{cd}=\left| \vec{r}_c - \vec{r}_d \right|$. 
The total mass is denoted by \be
	M := m_1 + m_2 + m_3 ~.
\label{def:M}
\ee

The system is invariant under translations and hence the total linear momentum \be
 \vec{P} := \vec{p}_1 + \vec{p}_2 + \vec{p}_3
\label{def:P}
\ee
 is conserved and the center of mass \be
 \vec{R}_{CM} := \frac{1}{m_1 + m_2 + m_3 } \sum_{c=1}^3 m_c\, \vec{r}_c \
 \label{def:R_CM}
 \ee
is in uniform motion. Therefore, we shall work in the center of mass frame, where the system has 6 degrees of freedom. The remaining conserved quantities are the total energy given by the value of the Hamiltonian (\ref{def:H}) and the total angular momentum \be 
	\vec{J} \( \{\vec{r}_c,\, \vec{p}_c \}_{c=1}^3 \) := \sum_{c=1}^3 \vec{r}_c \times \vec{p}_c ~.
\label{def:J}
\ee  
Their values are denoted by \be
 E, \vec{L} ~,
\ee
respectively. 

\presub {\bf Outcome parameters}. A decay into three free bodies is allowed only for positive $E$ trajectories. Negative $E$ trajectories, \be
 E < 0 ~,
\ee
can only decay into a single + binary and hence are presumably more long-lived and ergodic and will be assumed in this paper. 

\begin{figure}
\centering \noindent
\includegraphics[width=14cm]{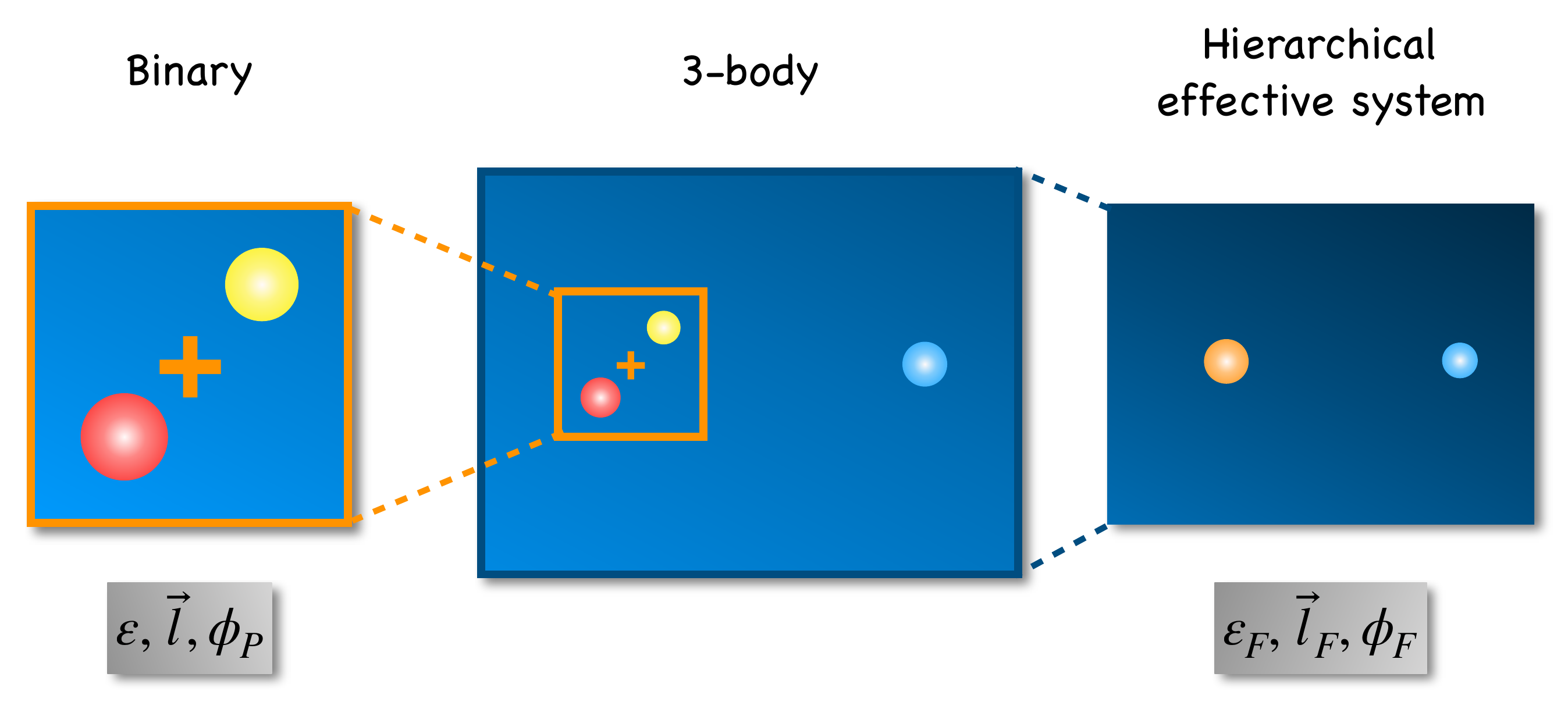} 
\caption[]{The system (middle) and its effective description. Left: the binary subsystem, together with its parameters. Right: the hierarchical effective system.}
 \label{fig1:coord_systems} \end{figure}

The final state of decay is an escaper, $m_s$, where $s$ is either $1,\, 2$ or $3$, moving away freely from a binary consisting of the masses $m_a,\, m_b$.  This motion decouples into the binary motion and the effective motion where the binary is replaced by an effective point particle, see fig. \ref{fig1:coord_systems}. 

For the Keplerian binary motion one defines \bea 
 m_B &:=& m_a + m_b  
 	\label{binary_mass} \\[7pt] 
 \mu_B &:=& \frac{m_a \, m_b}{m_B}  
 \label{def:mu_B}  \\[7pt]
\alpha_B &:=& G\, m_a \, m_b 
\label{def:alpha}
\eea
 which are the binary mass, binary reduced mass, and the potential strength constant. 

The binary's conserved quantities are \be
 \eps, ~ \vec{l}, ~ \phi_p ~, 
\ee
which denote the binary energy, angular momentum and the azimuthal angle for perihelion in the plane orthogonal to $\vec{l}$. $\eps,\, \vec{l}$ take values in the domain defined by the inequalities \bse
 \begin{align}
 \eps & \le E \label{binary_region:E_bdry} \\
-2\, \eps\, l^2 & \le k_s  \label{binary_region:circ_bdry} 
\end{align}
\label{binary_region}
\ese
where the binary constant is given by \be
 k := \mu_B\, \alpha_B^2 ~.
  \label{def:k} 
 \ee 
In the context of the 3-body problem, the identity of the escaper $s$ defines the binary under consideration and hence all binary parameters can display an $s$ index, e.g. $k_s$. 
Inequality \eqref{binary_region:circ_bdry} saturates for circular motion.

The effective motion replaces the binary by an effective point particle of mass $m_B$ located at the binary center of mass
\be
 \vec{r}_{B} := \frac{1}{m_B} \( m_a\, \vec{r}_a + m_b\, \vec{r}_b \)
\ee
The reduced effective motion describes a particle of reduced mass \be
 \mu_F = \frac{m_s\, m_B}{M} ~,
 \ee
position \be
 \vec{r}_F = \vec{r}_s- \vec{r}_{B} ~
 \label{def:rF}
\ee
 and an effective potential constant \be
 \alpha_F := G \, m_s \, m_B ~.
\ee
 The conserved quantities are \be 
   \eps_F, ~ \vec{l}_F, ~ \phi_F ~,
 \ee
 which denote the energy, the angular momentum vector and the azimuthal angle for perihelion within the plane orthogonal to $\vec{l}_F$. The effective $k$-constant is \be
k_F := \mu_F \, \alpha_F^2 ~.
\ee

If $\eps_F$ is positive and $\vec{r}_F$ is large enough than the effective motion is free and describes an escape. The effective motion is also useful for $\eps_F<0$ where is describes a sub-escape excursion.

\begin{figure}
\centering \noindent
\includegraphics[width=14cm]{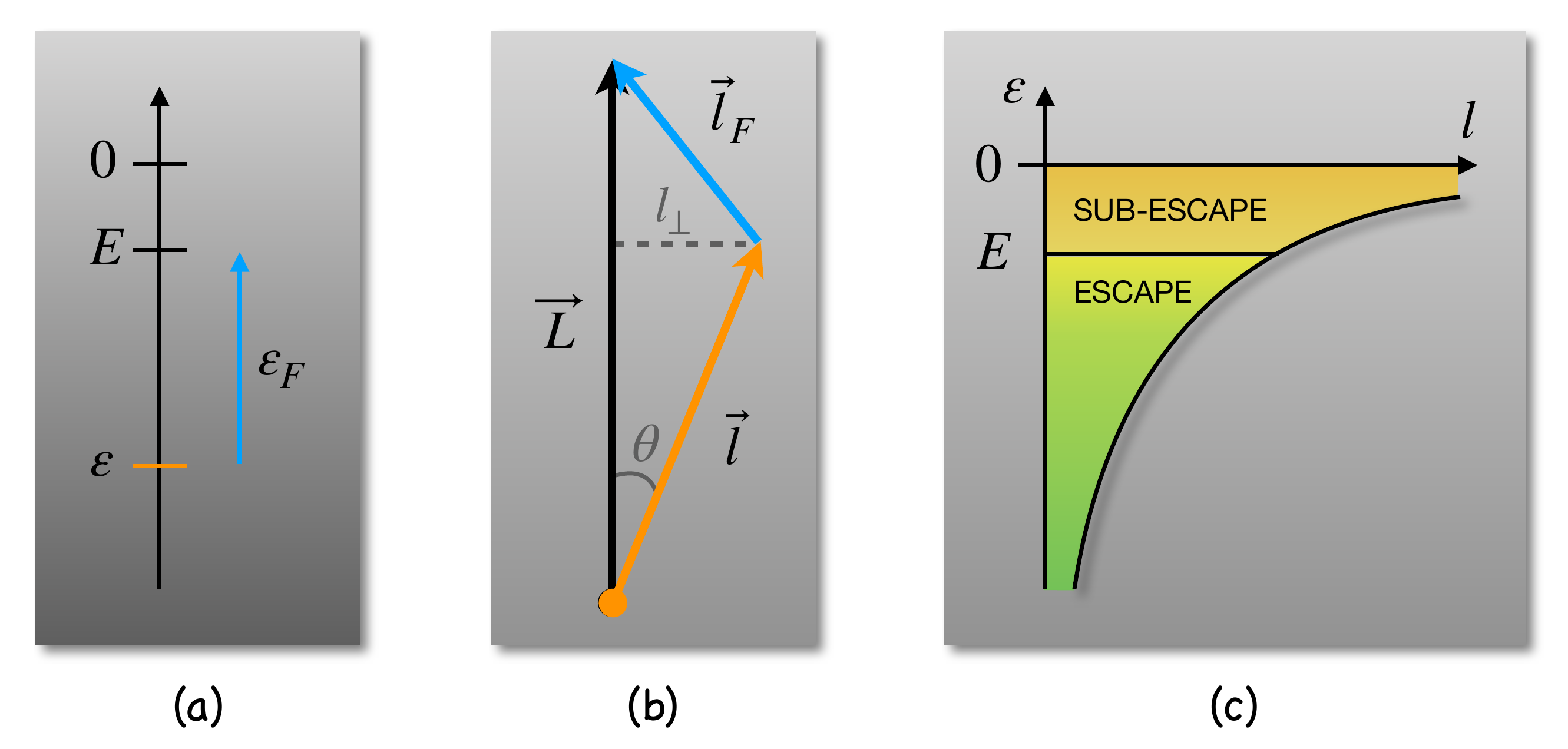} 
\caption[]{Energy and angular momentum space. (a) Conservation of energy. (b) Conservation of angular momentum. (c) The allowed region in the $\eps,\, l$ space  \eqref{binary_region:circ_bdry}.}
 \label{fig:E_L_space} \end{figure}

The conservation of energy and angular momentum and the $\eps,\, \vec{l}$ domain are illustrated in figure \ref{fig:E_L_space}.

\subsection{Presentation of the prediction} 
\label{subsec:prediction}

We turn to present the flux-based statistical prediction. Motivation and derivation will be given in section \ref{sec:derive}.

\presub {\bf Phase space volume flux}.  Consider the motion of a uniform density gas which fills the region in phase space with prescribed $E, \vec{L}$. The flux of phase-space volume (or phase-volume in short) is distributed over outcome parameters.  Due to time reversal symmetry the incoming and outgoing flux distributions coincide and need not be distinguished.  In the next section (after eq. \ref{flux_factorize}) it will be shown that the flux distribution is given by \be
 dF = 2 \pi\, dl_s(\eps)\; \frac{1}{l\, l_F} d^3 l\, d^3 l_F\,  \delta^{(3)}\(\vec{l}+\vec{l}_F-\vec{L}\)\; d\phi_F\, d\phi_p 
 \label{dF_infty}
\ee
 where $l_s$ is the maximum $l$ for given $\eps$, which is read from \eqref{binary_region:circ_bdry} and is given by \be
l_s(\eps) := \sqrt{\frac{k_s}{(-2 \eps)}} \qquad \textrm{ and hence } ~~ dl_s(\eps) = \frac{\sqrt{k_s}\; d\eps}{(-2 \eps)^{3/2}}
\label{def:ls}
 \ee
and where $\eps,\vec{l}$ belong to the domain (\ref{binary_region:circ_bdry}). 

\presub {\bf Chaotic absorptivity}. The phase-space of the three-body problem is known to be divided between regular and chaotic regions. 

The outcome states can also serve as initial conditions, so they can also be called asymptotic states. Given asymptotic state parameters $\eps_F, \, \vec{l}_F,\, \eps,\, \vec{l},\, \phi_p$ (due to symmetry there is no dependence on $\phi_F$) we define the chaotic subset \be
 \chi= \{ \psi_B \in [0,2 \pi] : \psi_B \textrm{ trajectory is chaotic} \} 
 \label{def:chi}
\ee 
 and the chaotic absorptivity \be
 \cE = \frac{1}{2 \pi}\, \mu(\chi)
 \label{def:cE}
 \ee  
 where $\psi_B$ is the binary phase which defines the initial state of the binary within its given orbit, more precisely, the so-called mean anomaly which is proportional to the time shift within the binary orbit, see \eqref{bound_cent_dsigma}, and where $\mu(\chi)$ denotes the measure (or mass) of $\chi$.  $\cE$ lies within the range $0 \le \cE \le 1$ and it describes the probability for an initial state of given parameters and with a random binary phase $\psi_B$ to evolve into an irregular, namely chaotic, trajectory.
 
By definition the phase-volume flux into the chaotic region is given by \be
dF_\chi =  \cE \, dF   \\
\label{dF_chi}
\ee
Thus, $dF_\chi$ factorizes into two factors: $dF$ which is given in closed-form in \eqref{dF_infty} and $\cE$ which is bounded. The conditions for $\cE$ to vanish will be studied in section \ref{sec:spec_pred}.  

\presub {\bf Ergodic approximation}. The ergodic approximation assigns equal probabilities to equal volumes in phase space, or equivalently, replaces time averages by ensemble averages over a fluid which fills phase space. This means that all ergodic trajectories have practically the same statistics, dictated by \eqref{dF_chi}, independent of their initial conditions. 

Subsection \ref{subsec:decay_rate} mentions numerical evidence implying that the chaotic region of the three-body problem is indeed ergodic to a good approximation. On the other hand, since the motion is not bounded, but rather has a finite decay time, the validity of the approximation is limited and would be interesting to study further. 

\presub {\bf Differential decay rate}. A decay rate is the probability per unit time to decay. A differential decay rate is a decay rate into outcomes which lie within a prescribed differential element of outcome parameters.  

In section \ref{sec:derive},  we shall motivate through the container analogy that within the ergodic approximation the differential decay rate is given by \be
 d\Gamma_s = \frac{dF_\chi}{\bsig_\chi}  
\label{dGm_s}
\ee
 where $dF_\chi$ is given by \eqref{dF_chi}, and $\bsig_\chi(E,\vec{L})$ denotes the regularized chaotic phase-space volume to be defined  in (\ref{def:reg_sigma_chi}).

\presub {\bf Regularized phase space volume}. The element of phase space volume (or phase-volume, in short) for a three-body problem with given $E,\, \vec{L}$ is given by \be
 d\sigma(E,\vec{L}) =    \prod_{c=1}^3 \( d^3 r_c\, d^3 p_c \) \;  \delta^{(3)}(\vec{R}_{CM}) \; \delta^{(3)}(\vec{P}) \;\delta^{(3)}(\vec{J}-\vec{L})\;  \delta( H-E) \;  ~,
\label{def:dsigma}
\ee
 where $\vec{R}_{CM},\, \vec{P},\, H,\, \vec{J}$ are functions of the phase space variables $\{\vec{r}_a,\vec{p}_a\}_{a=1}^3$ given at  (\ref{def:H} -- \ref{def:J}).  Its integration, \be 
  \sigma(E,L) = \int d\sigma(E,\vec{L}) ~ ~,
 \label{def:sigma}
  \ee 
  leads to a long distance divergence. This is regularized by choosing as reference the hierarchical effective description of the possible decay channels, consisting of one term for each possible escaper, through  \be
 \bsig(E,\vec{L}) = \sigma - \sum_{s=1}^3 \( \sigma ~ \backslash. ~ \{V \to V_s \} \) 
 \label{def:reg_sigma}
 \ee
where $a ~ \backslash. ~ r$ denotes the expression $a$ after performing a replacement according to rule $r$ (in the notation of the Mathematica computing system \cite{Mathematica}) and the hierarchical effective potential is given by \be
 V_s := - \frac{G\, m_a\, m_b}{r_{ab}} - \frac{G\, m_s\, m_B}{r_F}
\label{def:Vs}
\ee  
where $\vec{r}_F$ was defined in \eqref{def:rF} and it depends on $s$. In addition, the integration region for the reference is restricted to $\eps_F \ge 0$.  We note that in a 1d scattering problem the regularized phase-volume is nothing but the scattering delay time, as discussed around \eqref{scat_1d_sigma}. 

The chaotic regularized phase-volume is obtained by restricting the integration in \eqref{def:sigma} to the chaotic region \be
 \bsig \to \bsig_\chi
\label{def:reg_sigma_chi}
\ee
Since the regular trajectories are expected to have modest delay times, it might be reasonable to approximate $\bsig_\chi \simeq \bsig$

\presub {\bf Integration over momenta}. We developed a method to perform the momenta integrations in (\ref{def:dsigma}), see subsection \ref{subsec:method}, and we find  \be
 \sigma(E,\vec{L}) = 4 \pi\, \(\frac{\prod_c  m_c}{M}\)^{3/2} \int_{T_{\rm eff} \ge 0} \prod d^3 r_c\, \delta^{(3)}\(\vec{R}_{cm}\)\, \(\frac{2\, T_{\rm eff}}{\det I^{ij}} \)^{1/2}
 \label{sigma_red1}
\ee
where \bea
 I^{ij} &=& \sum_c m_c \( r_c^{\;2} \delta^{ij} - r_c^i\, r_c^j \) \non
T_{\rm eff} &:=& E - V - \half I^{-1}_{ij}\, L^i \, L^j ~.
\label{def:I_T}
\eea
 $I^{ij}$ is the moment of inertia tensor for a three body configuration and $T_{\rm eff}$, the effective kinetic energy, is the available kinetic energy after accounting for a centrifugal energy which is the minimal $T$ for momenta satisfying the conservation of angular momentum. 

The determination of $\bsig$ is the subject of work in progress where further analytic integration was found to be possible.

\begin{figure}
\centering \noindent
\includegraphics[width=14cm]{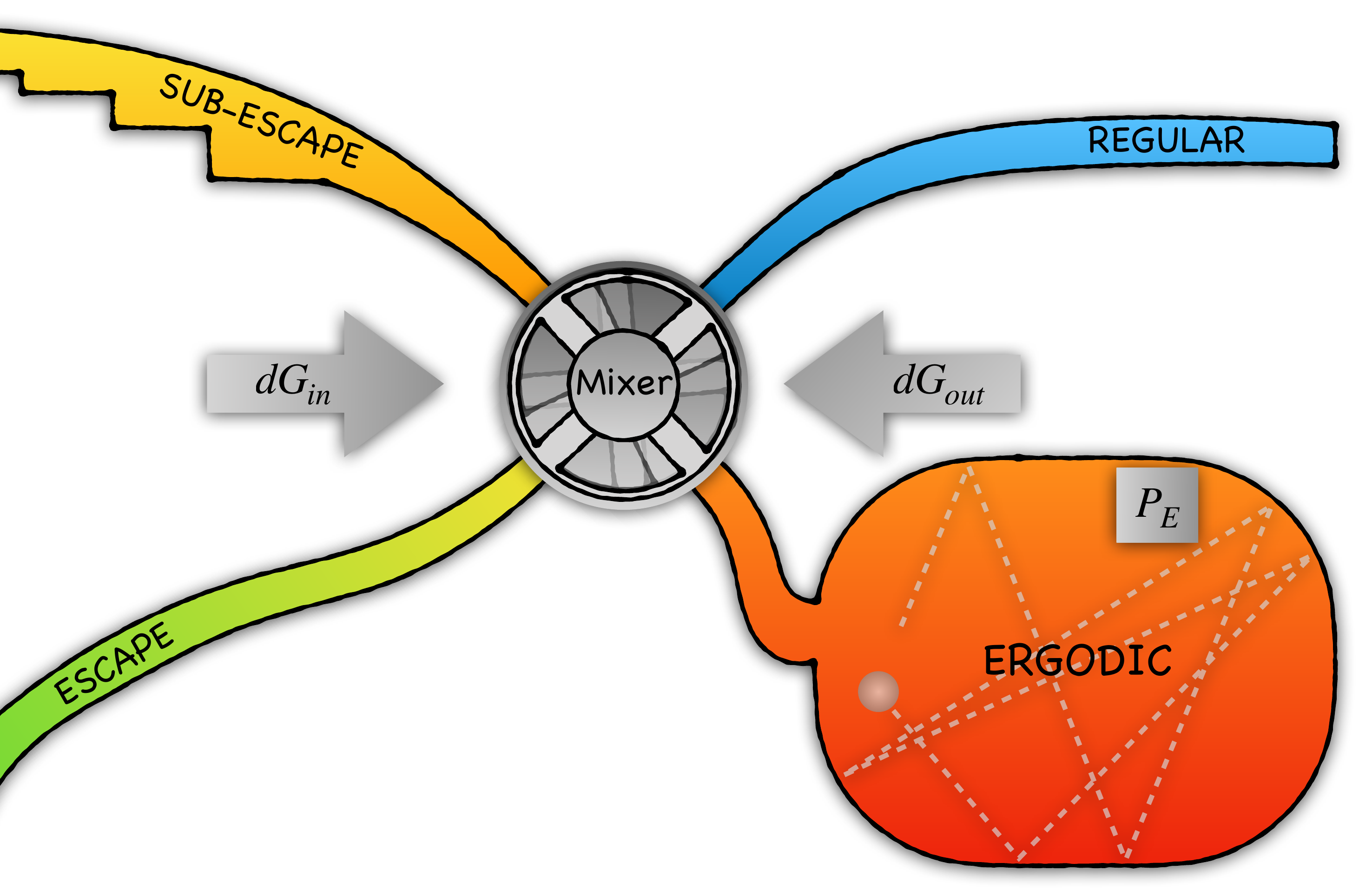} 
\caption[]{Piping diagram for the three-body system. E denotes the ergodic region, or compartment, while R denotes the regular region. The stay time in the E region is significantly longer than in the R region. Ejections are channeled through a mixer, or splitter, into either an escape or a sub-escape (excursion). Sub-escape durations are distributed over a range of possibilities. Upon re-entry they may reach either the E or R regions.}
 \label{fig:piping_diag} \end{figure}

\presub{\bf Statistical evolution}. In addition to ejections of a single body into an escape trajectory, the system also displays ejections into sub-escape excursions. These different population compartments can be represented by the piping diagram in figure \ref{fig:piping_diag}. 

We represent the different populations by the following time dependent variables \be 
	P_E,\, dG_{\rm out},\, dG_{\rm in}
\label{popul_var}
\ee
where $P_E$ represents the probability that the system is in the ergodic region; $dG_{\rm out}$ is the probability flux outgoing into hierarchical effective motion, distributed over outcome parameters which are denoted here collectively by $u$ \be
 u=\{ \eps, \, \vec{l}, \, \phi_p, \,  \eps_F, \, \vec{l}_F, \, \phi_F \} ~;
\label{def:u}
 \ee
  and $dG_{\rm in}$ is the probability flux incoming from the hierarchical effective motion. 

These variables satisfy the following system of statistical evolution equations \bea
 \dot{P}_E &=& -\Gamma_{\rm tot}\, P_E + \int \cE(u)\, dG_{\rm in}(u) \non 
 dG_{\rm out}(u) &=& d\Gamma(u) \, P_E + du \int \cE_R(u, v)\, dG_{\rm in}(v) \non
 dG_{\rm in}(u,t) &=& \begin{cases}
 dG_{{\rm in},\infty} (u,t)		& \eps_F \ge 0 \\
 dG_{\rm out}\(u', t-T\)		& \eps_F < 0
  \end{cases}
\label{evolution}
\eea
This equation system is explained in subsection \ref{subsec:stat_ev}. The notation is as follows. The total decay rate $\Gamma_{\rm tot}$ is given by \be
 \Gamma_{\rm tot} = \int d\Gamma = \int_{\eps_F \ge 0} d\Gamma + \int_{\eps_F < 0} d\Gamma ;
 \ee
 a dot denotes a derivative with respect to time $\dot{}=\delta/\delta t$ where we reserve the notation $d$ for differentials in the space of outcome parameters, e.g. $du$; $d\Gamma$ is the differential decay rate \eqref{dGm_s} only now its definition domain is extended to $\eps<0$ so as to include also the sub-escape ejections;  $\cE_R(u,v)$ is the mapping associated with absorption into regular motion; $dG_{{\rm in},\infty}$ is the asymptotic incoming probability flux, if any; $u'(u)$ is the excursion re-entry mapping and $T=T(u)= 2 \pi \sqrt{k_s}/(-2 \eps_F)^{3/2}$ is the excursion duration. 

The system requires as boundary conditions $P_E(0)$ as well as $dG_{\rm out}(u)$ for $-T(u) \le t \le 0$ and $dG_{{\rm in},\infty} (u,t)$ for $t \ge 0$. These are not standard ordinary differential equations with respect to time, due to the dependence on $t-T$, and may be termed ``a differential equation system with memory'' (alternatively, it can be recast as an integral equation).

By definition, a solution contains full information about the statistics of decay times.

This equation system does not account for the delay times and this is improved upon in subsection \ref{subsec:stat_ev}.  

\presub {\bf Numerical evidence}. Recently, outcome distributions were determined through numerical simulation for some initial conditions in \cite{MLT}. The escape probabilities were determined for 3 mass sets and they were compared with two predictions in table 12 there. The first prediction there is based on the current paper. The dependence on absorptivity is likely to cancel out in an averaged quantity such as the escape probability. Disregarding it as a first approximation,  the relevant  unnormalized probabilities are derived in section \ref{sec:unit_absrp}, and  according to  \eqref{x1_marg}, they are given by \be 
P_s =\frac{l_0^3}{3\,L} \propto k_s^3 ~.
\label{escape_prob}
\ee

The second prediction in \cite{MLT} is based on \cite{Stone_Leigh_2019}. Here we add a comparison to a third prediction which appears in the book \cite{Valtonen_book_2006}, eq. (7.23),  according to which the unnormalized probabilities are \footnote{
We note that \cite{Valtonen_book_2006} includes also the empirical expression $P_s =\frac{1}{m_s^q}$ with $q=3/(1+2 L^2/L_{max}^2),\, L_{max}=5/2 [(m_1 m_2 + m_2 m_3 + m_3 m_1)/3]^{5/4}/\sqrt{E}$, see equations (7.30, 7.33, 2.67, 7.28) there. We find it appropriate to compare our derived expression with a derived expression, rather than an empirical one.}
\be
P_s =\frac{1}{m_s^2} ~.
\ee

The numerical data and the three predictions are shown in table \ref{table:eject_prob}. It can be seen that prediction 1 is accurate at the 1\% level (absolute accuracy), while prediction 2 overshoots the probability inequality and prediction 3 undershoots. 

\begin{table*}
{\centering
\begin{tabular}{cccccc}
\hline
 Masses($M_\odot$) & Ejection Mass($M_\odot$) & Measured \cite{MLT} & Prediction 1 & Pred. 2 \cite{Stone_Leigh_2019}   & Pred. 3 \cite{Valtonen_book_2006} \\
\hline
& 15 (0) & $0.331(2) $  & 0.333 &  0.333 & 0.333 \\
15,15,15 & 15 (1) & $0.334(2)$ & \dittotikz  & \dittotikz & \dittotikz \\ 
 & 15 (2) & $0.335(2)$ & \dittotikz & \dittotikz & \dittotikz \\
\hline
 & 12.5  &$0.575(2)$  & 0.563 & 0.639 & 0.454\\
12.5,15,17.5 & 15  & $0.271(1)$ & 0.279 & 0.247 & 0.315\\
 & 17.5  & $0.154(1)$  & 0.158  & 0.114 & 0.231\\
\hline
 & 10  & $0.770(1)$  & 0.783 & 0.872  & 0.590 \\
10,15,20 & 15  & $0.167(1)$ & 0.159 & 0.103   & 0.262\\
 & 20 & $0.063(1)$ & 0.058 & 0.025 & 0.148 \\
\hline
\end{tabular}}
\caption{Ejection probabilities for ergodic trajectories measured in numerical simulations compared to the current prediction, as well as the predictions of \cite{Stone_Leigh_2019} and \cite{Valtonen_book_2006}. This table is based on table 12 of  \cite{MLT} where further information can be found.}
\label{table:eject_prob}
\end{table*}

It might be worth pointing out that agreement is reached even though each of the two methods, the statistical prediction and the numerical integration, requires altogether different calculations and concepts.

\newpage
\section{Derivation}
\label{sec:derive}

The previous section stated the statistical prediction for the outcomes and this section describes its derivation and the physical motivation behind it.

\subsection{Decay rate and phase-volume flux}
\label{subsec:container}

In the ergodic approximation the probability density $dP$ is taken to be proportional to the phase space volume density \be
 dP \propto d\sigma
\label{dP_sig}
 \ee
This relation was the basis for all previous studies of the statistical theory of the three-body system. However, once the system has decayed, its motion is free, and definitely not chaotic. Therefore the basic relation needs to be adapted. This becomes particularly necessary when one wishes to obtain cutoff independent predictions by sending the long distance cutoff (strong interaction radius) to infinity, thereby including more and more of the free motion phase space.  Moreover, while a probability distribution is the natural observable for bound motion, in cases where decay is possible the natural observable becomes the decay time.   

In order to formulate a more fitting theory, let us consider as a leading example a particle of energy $E$ moving in a container  of some arbitrary shape, which has perfectly elastically reflecting walls with a small hole. The escape of the particle out of the hole is analogous to a system decay: inside the container the motion is chaotic, while outside it is free. We are interested in the differential decay rate, namely the fractional probability of escape per unit time distributed over the exit direction.

From the perspective of Statistical Mechanics, the system's probability distribution can be considered to be a gas in phase space. According to the kinetic theory of gases the differential decay rate can be expressed  by \be
 d\Gamma = \frac{dF}{\sig} = \frac{d^3p\, \delta(H-E) \int d^3 r \, \delta(z-z_0)\, (\dot{z})_+}{\int d\sig} ~,
\label{dGm_dF_container}
 \ee
 where $dF=d^3p\, \delta(H-E) \int d^3 r \, \delta(z-z_0)\, (\dot{z})_+$ is the outgoing phase-volume flux through the hole distributed over outgoing momenta;  $z$ is a coordinate defined such that the hole is at $z=z_0$; $d\sig = d^3 r\, d^3 p\, \delta(H-E)$ is the micro-canonical distribution; the integration is over the hole's surface; $(x)_+$ denotes the ramp function defined by \be
 (x)_+ : = \begin{cases}
 x 	& 0 \le x \\
 0 	& x \le 0
\end{cases} ~\equiv~  \frac{x + |x|}{2} \equiv x\, \Theta(x)
\label{def:ramp}
\ee 
where $\Theta$ is the Heavyside $\Theta$-function; accordingly, $(\dot{z})_+$ is the outgoing normal velocity; and finally $\sig= \int d\sig$ is the total phase-volume within the container.
 
$d\Gamma$ has 1/time dimensions as is appropriate for a decay rate. The total decay rate is given by \be
	\Gamma = \int d\Gamma
\ee
 and the outcome probability distribution is \be
 dP = \frac{d\Gamma}{\Gamma} ~.
\label{dP}  
 \ee
$\Gamma^{-1}$ is proportional to the average decay time, $\tau_D$. The proportionality constant depends on the distribution of decay times: an exponential distribution, which is typical of random processes, implies \be 
\Gamma^{-1} = \tau_D ~,
\label{Gamma_tau2}
\ee
 while a uniform distribution implies $\Gamma^{-1}= 2 \tau_D$. $\tau_S$, the average scattering time, is related through \be
 \tau_S = 2\, \tau_D
 \label{tau3}
 \ee as a result of time reversal symmetry. 
 
Led by these examples we consider (\ref{dGm_dF_container}) to be the starting point of this paper, replacing (\ref{dP_sig}). However, since the three-body problem is unbounded, it may not be immediately clear where the location of the hole is, where the phase-volume flux $dF$ is defined, and why would $\sigma$ not be infinite. This will be answered in the following two subsections.

\subsection{Three-body phase-volume flux}

The determination of the three-body scattering flux will require us to first determine the two-body phase-volume and flux, which is the subject of the current subsection. We shall start with 1d motion and then turn to the central force problem, which is the essential part of the two-body problem. On the way, we shall build our understanding of the micro-canonical phase-volume, and the flux.
 
\presub {\bf 1d motion}. Let us denote the generalized coordinate by $x$, the momentum by $p$ and the Hamiltonian by \be 
	H=H(x,p)=\frac{p^2}{2m} + V(x) ~.
\ee The micro-canonical phase-volume element is given by \be
 d\sigma(E) := dx\, dp \, \delta(H-E) = dx\, \(\del H/\del p\)^{-1} = \frac{dx}{\dot{x}(x;E)} = dt
\label{dsigma_n_dt}
 \ee
 where in the second equality we have integrated over $p$ and the third equality uses Hamilton's equations. In words, the phase space element for fixed $E$ per $dx$ is that time element $dt$ of a trajectory with energy $E$ going through $x$ (either to the right or to the left).
 
For bound motion  (\ref{dsigma_n_dt}) implies \be
 \sigma(E) = T(E)
\label{bound_1d_sigma} 
\ee
where  $T(E)$ is the time-period, thereby reproducing $d\sigma/dE=T(E)$, see e.g. \cite{LL_mechanics}.

For scattering motion we require $\lim_{x \to \pm \infty}=0$ and $E>0$. The total time of a trajectory is infinite, however, a meaningful finite, or regularized, time can be defined, namely the scattering delay time \be
 T_{\rm delay}  := \int_{-\infty}^{+\infty} dx \(  \(\del H/\del p\)^{-1} -(2\, E/m)^{-1/2} \) \equiv \bsig(E)
\label{scat_1d_sigma}
\ee
In words, the delay time is defined by considering a particle arriving from $-\infty$ with energy $E$, moving under the influence of the potential $V(x)$, and a reference particle with the same initial conditions but free and unaffected by the potential. After the particle crosses the potential it returns to the initial velocity, but in general the two particle would be separated in position, which translates to a time delay (or advance).  On the other hand, according to (\ref{dsigma_n_dt}) $T_{\rm delay}$ can be interpreted as phase-volume regularized by subtracting a reference of free motion. These concepts were already discussed in \cite{Narnhofer_Thirring_1981}.

For $\sigma_{\rm reg} \equiv T_{\rm delay}$ to be finite $V(x)$ must tend to infinity faster than $1/x$. If it does not then the reference mechanics should retain the part of the potential which is of order $1/x$ or more, rather than free motion.

The phase-volume flux for scattering motion, defined in \eqref{dGm_dF_container}, is given by \be
 F(E) = \int d\sigma(E) \, \delta(x-x_0)\, \dot{x}(x) = 1
\label{scat_1d_F}
\ee
where the second equality uses \eqref{dsigma_n_dt}. Namely, the phase-volume flux of 1d motion is simply unity and is independent of the location $x_0$, where it is evaluated. 

\presub {\bf Central force}. Consider now a central force problem \be
 H=H(\vec{r},\vec{p}) = \frac{\vec{p}^{\;2}}{2 \mu} + V( \left| \vec{r} \right|) ~.
\ee

We start with two lemmas. \\
\noindent {\it Lemma 1}. Denote the direction of $\vec{J} := \vec{r} \times \vec{p}$ by the $z$ axis: $\vec{J}=J\, \hat{z}$. Then \be
 \int dz\, dp_z\, \delta(J_x)\, \delta(J_y) = \frac{1}{J} ~.
 \label{lemma1}
\ee
Proof: \bea
&& \int dz\, dp_z\, \delta(J_x)\, \delta(J_y) = \non
&=& \int dz\, dp_z\, \delta(p_x\, J_x)\, \delta(p_y\, J_y) \left| p_x \, p_y \right| = \non
&=& \int dz\, dp_z\, \delta(p_x\, J_x)\, \delta(- p_z\, J_z) \left| p_x \, p_y \right| = \non
&=& \frac{1}{\left| J_z \right|}  \int dz\,  \delta(-z\, p_x\, p_y) \, \left| p_x \, p_y \right| = \non
&=& \frac{1}{J} 
\label{proof:lemma1}
\eea 
where in passing to the third line we used $\delta(a) \delta (b) =\delta(a) \delta (b+a)$ and $p_x\, J_x + p_y\, J_y = - p_z\, J_z$, and in passing to the next line we integrated over $p_z$ and used $\left. p_x\, J_x \right|_{z=0} = -z\, p_x\, p_y$.  

\presub {\it Lemma 2}. The 2d planar phase space element transforms from Cartesian to polar coordinates as follows \be
 dx \, dy\, dp_x\, dp_y = dr\, d\phi\, dp_r\, dJ_z
 \label{lemma2} 
\ee 
Proof: in polar coordinates $dx \, dy = dr\, r d\phi$ while the momenta transform with the inverse Jacobian: $dp_x\, dp_y = dp_r\, dp_\phi/r$ and $p_\phi \equiv J_z$. Substituting these completes the proof. 

The Lemmas allow us to reduce micro-canonical phase-volumes for a central force into an effective radial problem as follows \bea
 d\sigma(E,\vec{L}) &:=& \int d^3r\, d^3p\, \delta(H-E)\, \delta^{(3)}(\vec{J}-\vec{L}) = \non
 &=&  \int d^2r\, d^2p\, \delta(H-E)\, \frac{\delta(J_z-L_z)}{\left| J_z \right|} = \non
 &=& \frac{d\phi}{L} \int dr\, dp_r\, \delta(H_{\rm eff}-E) = \non
 &=&  \frac{d\phi}{L}\,  dt_r
\label{bound_cent_dsigma}
\eea
where the first line is the definition of $d\sigma$ and in the central force context we distribute over both $E$ and $\vec{L}$; in passing to the second line uses lemma 1; the next uses lemma 2 and integration over $J_z$; $H_{\rm eff}:=p_r^2/(2 \mu) + L^2/(2 \mu\, r^2) + V(r)$ and the second term is called the centrifugal potential; and $dt_r$ is the time element for the radial problem. 

Applying our experience with 1d problems (\ref{bound_1d_sigma}--\ref{scat_1d_F}) to the radial problem we may immediately conclude that for bound motion \be
d \sigma(E,\vec{L}) = \frac{d\phi}{L}\, T_r(E) ~;
\label{bound_cent_sigma}
\ee 
while for a scattering motion we have \bea
 d\sigma(E,\vec{L}) & = & \frac{d\phi}{L}\, T_{\rm delay} \non
 \half\, T_{\rm delay} &:=& \int_{r_{\rm min}}^\infty dr \(\frac{1}{v_r(r;E)} - \frac{1}{\sqrt{2 E/m}} \) - \int_0^{r_{\rm min}} \frac{dr}{\sqrt{2 E/m}} ~.
 \label{scat_cent_sigma}
\eea
 Generalizing the 1d definition of the flux at infinity \eqref{dGm_dF_container}  we have\be	 
 dF(E,\vec{L}) :=  \lim_{R \to \infty} d\sigma \, \delta(r-R)\, \dot{r} = \frac{d\phi}{L} 		
	\label{scat_cent_F} ~.
 \ee

The determination of the phase-volume \eqref{bound_cent_sigma} and flux \eqref{scat_cent_F} of a central force problem can also be derived via the method to be introduced in subsection \ref{subsec:method} in a rather straightforward manner.

\vspace{.5cm}
Next we specialize the to two kinds of central force problems which are relevant to the gravitational three-body problem. It would be convenient to repackage $d\sigma,\, dF$ into $d{\tilde \sig}, d{\tilde F}$ defined by \bea
 d{\tilde \sig}(E,\vec{L}) &:=&  d\sig (E,\vec{L}) \, dE \, d^3 L \non
 d{\tilde F}(E,\vec{L}) &:=&  dF (E,\vec{L}) \, dE \, d^3 L
 \eea

\noindent {\bf Free 3d motion}. For free motion with reduced mass $\mu_F$, energy $\eps_F$ and angular momentum $\vec{l}_F$ equation (\ref{scat_cent_F}) implies that the asymptotic phase-volume flux is \be
 d{\tilde F}_F(\eps_F,\vec{l}_F) = d\phi_F \, d\eps_F \,  \frac{d^3 l_F}{l_F}
\label{dF}
\ee
where $\phi_F$ is the azimuthal angle for the escaper momentum in the plane orthogonal to $\vec{l}_F$.

\presub {\bf Kepler's problem}. Consider binary Keplerian motion as described in (\ref{binary_mass}--\ref{binary_region}). 
 The expression for the phase-volume element, (\ref{bound_cent_sigma}),  requires an expression for the period $T(E)$, which is obtained by a variant of Kepler's third law \be
T(E) = \frac{2 \pi}{\omega} = 2 \pi\,  \( \frac{a^3}{\alpha}\)^{1/2} = 2 \pi\,  \frac{\sqrt{k}}{(-2 \eps)^{3/2}}
\label{T_eps_relation}
\ee
where $a$ denotes the semi-major axis of the reduced motion and the binary constant $k$ was defined in (\ref{def:k}). Substituting back into (\ref{bound_cent_sigma}) we obtain \be
	d{\tilde \sigma}_B (E,\vec{l}) = 2\pi\, d\phi_p\, \sqrt{k} \frac{d\eps}{(-2 \eps)^{3/2}} \, \frac{d^3 l}{l} ~.
\label{dsigma_B}
\ee
 where $\phi_p$ is the azimuthal angle for perihelion in the plane orthogonal to $\vec{l}$. This result essentially appeared in \cite{Monaghan1_1976}, with references to the books \cite{Fowler1936,Jeans1929}.  

In fact, Keplerian motion has an additional conserved quantity which should be mentioned, namely the Laplace-Runge-Lenz (LRL) vector $\vec{A}$ which satisfies the algebraic relations \bea
 A^2 &=& \mu_B^2\, \alpha^2 + 2 \mu_B\, \eps\, l^2 
 \label{LRL_rel1} \\ 
0 &=& \vec{l} \cdot \vec{A}
 \label{LRL_rel2}
 \eea
These relations imply that $\vec{A}$ adds a single independent conserved quantity beyond $\eps,\vec{l}$ and we chose to parameterize it by $\phi_p$. Moreover, $A^2 \ge 0$ and (\ref{LRL_rel1}) imply the inequality (\ref{binary_region:circ_bdry}) in the $\eps,\, l$ plane. 

\presub {\bf Three-body flux}. In the three-body problem the outgoing phase-flux can be defined through the hierarchical effective motion  \be
  dF := \lim_{R \to \infty}  d\sigma \, \delta\(r_F-R\)\, \dot{r}_F
  \label{def:dFs}
 \ee
where $\vec{r}_F$, the relative position of free motion, was defined in (\ref{def:rF}).  

As discussed in the previous section, the asymptotic states for negative energy three-body motion consist of a widely separated binary and a free body. Accordingly, the phase-volume flux into asymptotic states factorizes as \be
 dF = \int d{\tilde F}_F\; d{\tilde \sigma}_B\; \delta(\eps+\eps_F-E) \, \delta^{(3)}\(\vec{l}+\vec{l}_F-\vec{L}\)
 \label{flux_factorize}
 \ee
where $d{\tilde F}_F$ denotes the flux for free motion and $d{\tilde \sigma}_B$ denotes the phase-volume of binary Keplerian motion. 

Substituting (\ref{dF},\ref{dsigma_B}) into (\ref{flux_factorize}), we complete the derivation of \eqref{dF_infty}.
 
We note that it  depends on parameters only through the relevant binary constant $k_s$ (\ref{def:k}) which appears in both the period--energy relation (\ref{T_eps_relation}) and the $\eps,\, l$ relation for circular motion, which sets a boundary for the allowed values of these quantities (\ref{binary_region:circ_bdry}).

\subsection{Three-body decay rate}
\label{subsec:decay_rate}

The general relation between decay rate and outgoing flux \eqref{dGm_dF_container} relied on the ergodic nature of the particle's motion inside in the container.
Simulations provide some evidence that the chaotic region of the three-body problem is ergodic as well. First, decay times are known to approximately obey an exponential distribution, which are the signature of a random process. This is known at least since \cite{Valtonen_1988} and was demonstrated again recently in \cite{MLT}. Secondly, \cite{MLT} finds that in the equal mass case the escape probability is approximately equal despite markedly different initial conditions. Independence of initial conditions is another signature of an ergodic motion. Finally, comparison with simulations (see table \ref{table:eject_prob}) are also supportive of this conclusion. 

We would like to define a differential decay rate for the three-body problem in analogy with the case of the container.
Since the system is unbounded, the previous subsection suggests that the definition (\ref{dGm_dF_container})  should be generalized by taking the flux from the chaotic region $dF_\chi$ and the regularized the phase-volume of the chaotic region, $\bsig_\chi$, namely \be
	d\Gamma = \frac{dF_\chi}{\bsig_\chi} ~.
 \label{dGm_dF}
\ee
The chaotic absorptivity relates $dF_\chi$ with the total flux $dF$ through \eqref{dF_chi}, and $dF$ was derived to be \eqref{dF_infty}.  We note that $dF$ depends only on the asymptotic states which are always simpler than the full interacting system.

Decay and scattering times are related, see \eqref{tau3}, and hence chaotic decay is related to chaotic scattering
which is a well-studied theory, see e.g. \cite{Gaspard_book_1998,Ott_Tel_1993,Seoane_Sanjuan_2013}. The author expects that the relation (\ref{dGm_dF}) is known, but so far was unable to find the reference for its introduction and would appreciate correspondence on this matter. For a good numerical study of the three-body problem from the perspective of chaotic scattering see \cite{Boyd_McMillan_1993}. 

\presub {\bf Regularization of phase-volume}. In this part we motivate the expression for the regularized phase-volume of the three body system \be
 \bsig = \bsig (E,L;\; m_1, m_2, m_3) ~.
 \ee
 
 The unregularized phase-volume is given by (\ref{def:dsigma},\ref{def:sigma}). Configurations with two nearby masses which are widely separated from the third can have $V < E$ and hence contribute to $\sig$. Moreover, since the separation between the binary and the escaper can be arbitrary, the contribution from these configuration diverges. 
 
We have already encountered a similar divergence in 1d, where $\sig$ was regularized in \eqref{scat_1d_sigma} by setting the asymptotic states as reference and subtracting them from the integrand. There are three kinds of asymptotic states and accordingly, the regularization should take the form \be
 \bsig = \sig - \sum_{s=1}^3 \sig_s
 \ee
 
$\sig_s$ should describe the available phase-volume to asymptotic states, once the potential energy between the binary constituents and the single are turned off. The binary potential energy should certainly be retained. Moreover, we can sill include an interaction term between the single and the binary's center of mass.  If fact, it is necessary to include the relevant gravitational potential in order to regularize the long-distance Kepler divergence mentioned below \eqref{scat_1d_sigma}. Altogether we take \be
 \sig_s = \sig ~ \backslash. ~ \{V \to V_s \}
 \ee
 where $V_s$ was defined in \eqref{def:Vs}. This completes the motivation behind the the definition \eqref{def:reg_sigma}.

We comment that we have seen already in (\ref{scat_1d_sigma}) that the 1d regularized $\sigma$ is related to the decay time. Similarly, we may think of the regularized three-body phase-volume as an averaged decay time.

\subsection{Determination of phase-volume through conjugate variables} 
\label{subsec:method}

In this subsection we present a method to perform phase space integrations over the momenta.

The definition of phase space volume  (\ref{def:dsigma},\ref{def:sigma}) can be written as \bea
 \sigma &=& \int  \( \prod_c  d^3 r_c \) \;   \delta^{(3)}(\vec{R}_{CM}) \; \rho \non
  \rho &=& \rho(\{\vec{r}_c\}):= \int  \( \prod_c d^3 p_c \) \;   \delta^{(3)}(\vec{P}) \; \delta( H-E) \; \delta^{(3)}(\vec{J}-\vec{L}) ~.
\label{def:sigma_hat}
\eea

The expression for $\rho$ can be interpreted geometrically as follows. The positions $\vec{r}_1, \, \vec{r}_2, \, \vec{r}_3, $ define a plane. If we denote the orthogonal direction by $z$ then the momenta components transverse to to the plane, $p_{z1}, p_{z2}, p_{z3}$, satisfy three linear constraints corresponding to $J_x, J_y, P_z$. The constraints are independent as long as the positions are not collinear and hence these transverse momenta are uniquely fixed. These equations are analogous to those in the statics problem of the three-legged table, where the three normal forces are unknown. The planar momenta components are constrained by the Hamiltonian to lie on a 5-ellipsoid. The $P_x, P_y$ constraints intersect this ellipsoid with a codimension-2 plane passing through the origin thereby defining a 3-ellipsoid. Finally the $J_z$ constraint intersects this ellipsoid with a codimension-1 plane passing in general outside of the origin (analogous to a non-zero latitude), thereby defining a 2-ellipsoid to integrate over.  

This geometrical approach should allow to perform the integration, but we did not take this way. Instead we choose to express the $\delta$-functions through the well known identity \be
 \delta (t) =  \int d\nu \exp 2 \pi i\, \nu\, t ~.
 \label{delta_id}
 \ee
Note that we used a symmetrical convention for the $2 \pi$ factors, which differs from the more popular convention in the physics literature. The idea is to lift the argument of the $\delta$-functions into an exponential function. Since all of these arguments depend on the momenta $\vec{p}_c$ at most quadratically, the momentum integrations will reduce to Gaussian integrals. In this sense, this method resembles  Schwinger parameters in the evaluation of Feynman diagrams. 

More concretely, for the three-body problem we use (\ref{delta_id}) to replace \bea
  \delta( H-E)		&=& \int d\beta  \exp 2 \pi i\, \beta \, (H-E) \non
 \delta^{(3)}(\vec{P}) &=& \int |\beta|^3\, d^3 U \exp \(-2 \pi i\, \beta\, \vec{U} \cdot \vec{P} \) \non
 \delta^{(3)}(\vec{J}-\vec{L}) &=&  \int |\beta|^3\, d^3 \Omega \exp \( -2 \pi i\, \beta\, \vec{\Omega} \cdot (\vec{J}-\vec{L}) \)\non
  \eea
Namely, we denote the by $\beta$ the variable conjugate to energy (this $\beta$ is \emph{not} the same as the inverse temperature which appears in the canonical ensemble, but there is some similarity); the variables conjugate to $\vec{P}$ is denoted by $\beta \vec{U}$ where $\vec{U}$ can be thought to represent a velocity of the center of mass; and finally the variables conjugate to $\vec{J}$ is chosen as $\beta \vec{\Omega}$ where $\vec{\Omega}$ can be thought to represent an angular velocity.

Combining these replacements into (\ref{def:sigma_hat})  we obtain \bea
 \rho &=& \int_{-\infty}^{+\infty} d\beta\, |\beta|^6 \, d^3 U\, d^3 \Omega \, \( \prod_c d^3p_c \)\exp 2 \pi i\, \beta (H_1-E) \non
 H_1 &:=& T + V - \vec{U} \cdot \vec{P} - \vec{\Omega} \cdot (\vec{J}-\vec{L}) ~.
\eea  

Since $\vec{p}_c$ appears quadratically in $T$ and linearly in $\vec{P}, \, \vec{J}$ the integration can be performed by the complex Gaussian integral \be
 \int_{-\infty}^{+\infty} dx\, \exp \pm \pi i\, x^2 =  \exp \pm \frac{2 \pi i}{8}
 \ee
derived by deforming the integration contour to $x= \exp (\pm 2\pi i\,/8) \; y $ where $y$ is real.

The result of the Gaussian integration over $\prod_c d^3 p_c$ is the determinant \be
D_p = \( \frac{ (\prod_c m_c)^3}{\left| \beta \right|^9} \)^{1/2} \exp 2 \pi i\, \frac{9}{8} \sgn \beta
\ee and eliminating $\vec{p}_c$ from $H_1$ (or equivalently, completing the square) $H_1$ is replaced by \be
 H_2 = V - \frac{M}{2} U^2 - \half I^{ij} \, \Omega_i \, \Omega_j + \vec{\Omega} \cdot \vec{L} 
\ee 
 where $M, \, I^{ij}$ were defined in (\ref{def:M},\ref{def:I_T}). 
In deriving $H_2$ we set $\vec{R}_{cm}=0$ due to the $\delta$-function in (\ref{def:sigma_hat}).

Now the dependence on $U$ is quadratic and hence the $d^3 U$ integration can be performed leading to \bea
 D_U &=& \( \frac{ 1}{M^3\, \left| \beta \right|^3} \)^{1/2} \exp \(-2 \pi i\, \frac{3}{8} \sgn \beta \) \non 
 H_3 &=& V - \half I^{ij} \, \Omega_i \, \Omega_j + \vec{\Omega} \cdot \vec{L} ~.
 \eea
 
The dependence on $\Omega$ is quadratic as well and hence the $d^3 \Omega$ integration can be performed to give \bea
 D_\Omega &=& \frac{ 1}{\(\det I^{ij} \left| \beta \right|^3\)^{1/2}} \exp \(-2 \pi i\, \frac{3}{8} \sgn \beta \) \non 
 V_{\rm eff} &=& V + \half I^{-1}_{~ij} \,L^i\, L^j  ~,
 \eea
where $H_3$ is replaced by $V_{\rm eff}$, so denoted since it generalizes the effective potential of the two-body problem by adding a centrifugal potential which accounts for the minimal kinetic energy which must arise in the presence of the $\vec{J}$ constraint. 

The final integration over  is  \bea
 & & 2 \operatorname{Re} \int_0^\infty \frac{d\beta}{\beta^{3/2}} \, \exp 2 \pi i \frac{3}{8} \; \exp \( - 2 \pi i\, \beta \, T_{\rm eff} \) =  4 \pi  \operatorname{Re} \sqrt{2 T_{\rm eff}} \non  
  T_{\rm eff} &:=& E - V_{\rm eff}
 \label{beta_int_reslt}
\eea
and $T_{\rm eff}$ denotes the available kinetic energy. The integral can be evaluated by setting $n=3$ in the following complex $\Gamma$-type integral \be
 \int_0^{\infty} \frac{d\beta}{\beta^{n/2}} \, \exp (\pm  i\, \beta)  =  \exp \( \pm 2 \pi i\frac{2-n}{8}\) \; \Gamma\( \frac{2-n}{2} \)
 \ee
which is derived by deforming the integration contour to $\beta= \pm i\, \hat{\beta}$ where $\hat{\beta}$ is real. The $4 \pi$ factor in (\ref{beta_int_reslt}) is related to the area of the 2-ellipsoid from the geometrical interpretation. 

Combining all integrations we obtain \be
 \rho = 4 \pi\, \[ \(\frac{\prod_c  m_c}{M}\)^3 \frac{2\, T_{\rm eff}}{\det I^{ij}} \]^{1/2}
 \ee
 within the domain $T_{\rm eff} \ge 0$ thereby completing the derivation of \eqref{sigma_red1}.  We note that the factor $\prod_c  m_c/M$ restricts in the two-body problem to the reduced mass (\ref{def:mu_B}). Further reduction and evaluation of $\sig$ are currently in progress.

Summarizing, we presented a method to perform the momentum integration for the phase space volume integral of the three-body problem, where each $\delta$-function is represented as an integral over an auxiliary conjugate variable. Clearly, the method is general and applies to any micro-canonical phase space integration where a $\delta$-function depends quadratically on some integration variable. In particular, it immediately generalizes to the $N$-body case, including the constraints for the conservation of total linear momentum and total angular momentum. 

This method associates a phase with any point in phase space and then integrates over all phases. In this way it is related to the thermal partition function, where each point is assigned a Boltzmann factor (the two methods are related in the same way that a Fourier transform is related to a Laplace transform). At the same time, the sum over phases connects with path integral, yet here one integrates over phase space and not over paths. In the language of probability theory this would be a characteristic function of the energy distribution \cite{Charac_wiki}.

We found it instructive to use this method to calculate the area of an $n$-sphere, the phase-volume of a two-body problem and that of a central force.  

The author expects that this method is known, but was so far unable to find it in the literature and would appreciate correspondence on this matter.

\subsection{Statistical evolution}
\label{subsec:stat_ev}

This subsection addresses the distribution of decay times. The distribution is known to be approximately exponential.  This distribution appeared in the review \cite{Valtonen_1988} while the use of the half-life term appeared already in \cite{Valtonen_1974,Valtonen_Aarseth_1977}. However,  \cite{Agekian_Anosova_Orlov_1983} argued that the average lifetime is infinite due to long sub-escape excursions and this is not consistent with an exact exponential distribution. A late time correction to the exponential distribution, the heavy tail power law, was described in \cite{Shevchenko_2010} and detected by simulation shortly after in \cite{Orlov_Rubinov_Shevchenko_2010}.  Essentially, long sub-escape excursions are responsible for the power law tail.

In order to model the distribution of decay times, we should understand the time evolution of a typical chaotic initial state. Some time after the motion begins, gravitational attraction selects two bodies that rush towards each other and undergo a close encounter, which can be summarized as a sort of elastic collision. Next, the observing body selects one of the bodies which emerges from previous encounter and a second close encounter unfolds. After several rounds of close encounters an outgoing body may have enough energy as to be ejected, leaving behind a binary. If the total effective energy is negative, namely $\eps_F < 0$, the ejected particle will turn back at some point, moving on an eccentric ellipse. This is a sub-escape excursion. Next it re-approaches the binary with the result of either a return to a chaotic state, with probability $\cE$, or a regular motion resulting in another ejection, not necessarily of the same body. The motion alternates between chaotic close encounters and sub-escape excursions until it ends in an escape, that is, an ejection event with  $\eps_F > 0$.

In order to model this system we construct a sort of population dynamics based on the piping diagram in figure \ref{fig:piping_diag}. The notations are defined in  \eqref{popul_var} and the associated equation system is given in \eqref{evolution}.

We comment that probability fluxes are denoted by $G$, while the more general phase-volume fluxes  are denoted by $F$. The $G$ fluxes describe more special ensembles which are  determined by initial conditions. We mention that the three-body problem is known to exhibit also hierarchical long-lived trajectories, not mentioned thus far, but it is so far unclear whether those are significantly coupled to the ergodic region.

Let us discuss the equation system  \eqref{evolution}. The first equation expresses that the ergodic region loses flux to the outgoing flux into hierarchical motion, either escape or sub-escape, and gains flux from the absorptive part of the incoming flux. The second equation describes the two contributions to the outgoing hierarchical flux: the first term describes the flux leaving the ergodic region and already described in the first equation while the second term describes the redistribution of flux which undergoes regular motion and is characterized by the regular motion mapping $\cE_R$. The final equation describes the sources of the incoming hierarchical flux: the first contribution is from asymptotic incoming boundary conditions while the second contribution is from the return from sub-escape excursions. 

The regular motion mapping $\cE_R(u,v)$ satisfies conservation of probability \be
 \int du\, \cE_R(u,v) = \cE_R(v) :=1-\cE (v)  
\label{calE_R_prob}
 \ee
and compatibility with the phase-space filling flow \be
 \int \cE_R(u,v) \, dF(v) = \cE_R(u)\, dF(u) 
\label{calE_R_stab}
\ee
where $dF$  is the asymptotic effective flux distribution defined in (\ref{dF_infty}). 

This equation system satisfies two tests. First, as expected, it conserves the total probability. The latter is given by \be
P_{\rm tot} = P_E + \int \int_{t-T}^t \delta t' \, dG_{\rm out}(t') + \int_{t}^\infty \delta t' \,  \int dG_{{\rm in},\infty}(t')
 \ee
where \eqref{calE_R_prob} was used. Given that the system is linear, $N_{\rm tot}$ can always be normalized to 1.

Secondly, the flow describing a full phase-space should be a solution. This is obtained once one requires \eqref{calE_R_stab}.

An initially ergodic state is described by the initial condition $P_E(0)=1$ while the initial fluxes vanish.  The corresponding solution to the equation system \eqref{evolution} describes both the process of escape and that of sub-escape excursion and hence I believe it will display both the initial exponential distribution and the power law tail.

\presub {\bf Statistical evolution with delay time}. This equation system does not account for the delay time in the process of absorption into either the ergodic or the regular region. This is improved upon by the following equation system.  

The equation system \eqref{evolution} can be improved upon by generalizing $\cE(u) \to \cE(u,t)\,  \delta t$ which denotes the absorption fraction whose delay time lies between $t$ and $t + \delta t$. Similarly, we generalize $\cE_R(u,v) \to \cE_R(u,v,t)\,  \delta t$. The resulting equation system is \bea
 \dot{P}_E &=& -\Gamma_{\rm tot}\, P_E + \int^t \delta t' \int \cE(u,t-t')\, dG_{\rm in}(u,t') \non 
 dG_{\rm out}(u) &=& d\Gamma(u) \, P_E + du \int^t \delta t' \int \cE_R(u, v,t-t')\, dG_{\rm in}(v,t') \non
 dG_{\rm in}(u,t) &=& \begin{cases}
 dG_{{\rm in},\infty} (u,t)		& \eps_F \ge 0 \\
 dG_{\rm out}\(u', t-T\)		& \eps_F < 0
  \end{cases} ~.
  \label{evolution2}
\eea

\newpage
\section{More predictions}
\label{sec:spec_pred}

In this section we provide additional analysis and  predictions, including the dimensionless parameters, the gross features of the absorptivity, and a prediction for escapes by a narrow margin.

\presub {\bf Dimensionless parameters and limits}.   The decay rate distribution (\ref{dGm_s}) depends on the following parameters: the conserved quantities $E,\vec{L}$, the masses $m_1, m_2, m_3$ and $G$. We define the following 3 dimensionless parameters \be
 x_s^2 := \frac{-2 E\, L^2}{k_s}
\label{def:xs}
\ee
These dimensionless quantities are independent and complete (any other dimensionless quantity must be a function of these).

The problem has two limits in which the problem simplifies \bi
\item $L=0$ and hence $x_s=0, ~s=1,2,3$
 \item $x_s \ge 1$. 
 Strictly speaking this $x_s \ge 1$ is a region in parameter space, but from the perspective of the interval $0 \le x_s \le 1$, $x_s=1$ is a limit. 
\ei 

The $\vec{L}=0$ limit displays an enhancement of conserved quantities. A generic configuration of the three bodies defines a plane. Denoting this plane by $x,y$ coordinates, one finds that $\{p_{z,c}\}_{c=1}^3$, the orthogonal components of the momenta, satisfy 3 constraints associated with $P_z,\, J_x,\, J_y$. For $\vec{L}=0$ these constraint are homogeneous and hence they force $p_{z,c}=0$ for all bodies. This means that the motion will remain limited to the original plane and that $\{p_{z,c}\}_{c=1}^3$ are conserved quantities. Note that the above-mentioned system of three equations for three unknowns is completely analogous to (the homogeneous part of) the equation system in the statics problem of the three-legged table (for the three normal forces). In the degenerate case of an initially collinear configuration one can show that the motion is still planar.  

High $L^2$ suggests that the impact parameter for the free motion will be significantly larger than the binary orbits size and hence a regular motion would be more likely than a chaotic one, see also around \eqref{flyby_certain} below. 


\presub {\bf Model of absorptivity}. The chaotic absorptivity, defined in \eqref{def:cE} lies in the range \be
 0 \le \cE \le 1 . 
\ee
It is interesting to find the domain where $\cE=0$ because outside (a neighborhood) of it, the (logarithm of the) absorptivity would be bounded, and hence the differential decay rate, $d\Gamma$, would be bounded as well, according to  (\ref{dF_chi}-\ref{dGm_s}). 

The vanishing of $\cE$ means that the motion is regular with certainty. There are several known types of regular motion: the flyby, where the original binary survives the scattering; the exchange, where the incomer undergoes a close encounter with one of the binary components which gets ejected; and more. A weak enough scattering guarantees with $\cE=0$. This happens whenever \be
 \rho := \frac{R_{{\rm min},F}}{R_{{\rm max},B}} \ge \rho_c \sim 1
\label{flyby_certain}
\ee
where $R_{{\rm min},F}$ is the periastron for the effective motion of the single body with respect to the binary, namely, the minimal distance between the single and the binary,  $R_{{\rm min},F}$ is the apastron of a binary component relative to the the binary center of mass, and $\rho_c$ is a critical value of order $1$ which depends on the  outcome parameters. 

Moreover, it appears reasonable that this domain of flyby certainty is the only place where $\cE$ vanishes.

For $\rho \le \rho_c$ we have a phase transition into $\cE>0$, and for small sub-critical $\rho$ a flyby critical exponent $p_f$ is expected \be
 \cE \simeq (\rho_c-\rho)_+^{p_f}
 \label{def:p_f}
 \ee
where $(x)_+$ denotes the ramp function defined in \eqref{def:ramp}. Moreover, on general grounds $p_f$ should be determined by a linearized analysis of a critical flyby trajectory.  


\presub{\bf Tight enough binaries do not absorb}. Let us see that for $L>0$ tight enough binaries imply vanishing chaotic absorptivity. 

Given $L>0$ the following minimal binding energy $(-2 \eps) > k_s/L^2$ guarantees that  $l_F$ is bounded away from 0: $l_F \ge L-l_s(\eps) > 0$ 
where $l_s(\eps)$ was defined in \eqref{def:ls}. Furthermore, for large enough $\eps_F$ (when the $e_F$ the eccentricity for the effective motion is in the range $e_F \gg 1$) $R_{{\rm min},F}$ can be approximated by \be
 R_{{\rm min},F} \simeq \frac{l_F}{\sqrt{2 \mu_F\, \eps_F}} ~.
 \ee
 
At the same time  $R_{{\rm max},B}$ can be bounded from above by  \be
 R_{{\rm max},B} = \frac{\max\{m_a,m_b\}}{m_B}\, (a+c) \le 2a = \frac{\alpha_F}{(-\eps)}
\ee 

Combining the last two relations we get \be
 \rho^2 \ge \frac{\( L- l(\eps) \)^2}{2 \mu_F\, \alpha_B^2\, }\, \frac{\eps_B^2}{\eps_F}
\ee 
By increasing the minimal binding $\eps_B^2/\eps_F \simeq \eps_F$ can be increased so that the RHS is $\ge \rho_c^{~2}$ which would imply certain flyby \eqref{flyby_certain}. This completes the argument.


\presub{\bf Prediction for escapes by a narrow margin}. For small values of $l_F$ the flyby fraction decreases \eqref{flyby_certain} and the distribution would be expected to be somewhat independent of $\cE$. In fact, for $l_F=0$ the flyby fraction vanishes and the system always experiences an irreducible three-body episode. However, other forms of regular motion are still possible, including exchange, see e.g. \cite{MLT}.

Having in mind the simulations of \cite{MLT}, we shall assume the $x_s \ge 1$ limit. 
 In this case
 there exists a threshold (lower bound) for $l_F$, namely \be
 l_F \ge l_{F,c} := L-l_0 >0
\ee
where \be
l_0 := l_s(E) \equiv \sqrt{\frac{k_s}{(-2\, E)}} ~.
\label{def:l0}
\ee

For $l_F$ just above threshold we expect the distribution to be independent of $\cE$. Therefore we consider the unnormalized outcome distribution \be
dP_s = dF_s
\ee
where $dF_s$ is the flux distribution \eqref{dF_infty}, approximating $\cE$ to be constant. Integrating over $\eps$ and the angular variables of $\vec{l}_F$ we find  \be
dP_s \propto (l_F-l_{F,c})_+^2 \, dl_F/l_0^{~3} ~.
\label{lf_distrib}
\ee
This means that the a critical exponent with value 2 is predicted just above the $l_F$ threshold.

\section{Discussion}
\label{sec:discussion}

The main result of this paper is a statistical prediction for the outcome of three-body motion. Fig. \ref{fig:piping_diag} provides a schematic description of the system, and the associated equation system of statistical evolution appears in \eqref{evolution}. The differential decay rate, an essential ingredient of the equation system, is reduced in an exact way into three factors  \eqref{dGm_s} \bi    
 \item The flux, given in closed-form in \eqref{dF_infty},
 \item The absorptivity, defined in \eqref{def:cE}, which is bounded to lie in the range $0 \le \cE \le 1$,
 \item The regularized phase-volume, defined in (\ref{def:dsigma}-\ref{def:reg_sigma_chi}), which serves as a normalization. 
 \ei

The prediction provides both the outcome statistics and the decay time statistics. Unlike previous treatments, the analysis is based on the notion of the flux, it is cutoff independent, it does not contain a spurious parameter, and Delaunay elements were not used. Moreover, it significantly improves the agreement with numerical data.
 
\presub {\bf Future research}. The predictions of section \ref{sec:spec_pred} are in the process of being tested against statistics from a large number of numerically integrated trajectories. The results will hopefully appear in a companion paper. So far the simulated results show good agreement with (\ref{escape_prob},\ref{lf_distrib}) and outperform previous approaches. It might be worth pointing out that agreement is reached even though each of the two methods, the statistical prediction and the numerical integration, requires altogether different calculations and concepts.

A study of $\bsig$ is in progress. As a result of the above-mentioned reduction, the problem is reduced to the study of $\cE$, which is bounded and presumably simpler. 

This study assumes the ergodic approximation which is necessarily imperfect given that the system is open. It would be interesting to quantify the associated error and, if relevant, to develop corrections to it.

Possible implications for other problems have not been overlooked. In particular, this work should easily generalize to the electrostatic three-body problem. In that case negative total energy implies that the charges cannot all have the same sign, and the escaper can only be one of the two identical sign bodies. 
 
\vspace{0.5cm}
 
In conclusion, the three-body problem is one of the richest and longest standing problems in physics. Since Poincar\'e it was believed to be unsolvable as long-term predictions are impossible due to its chaotic nature. However, a prediction in probability is possible, and the author believes that this work is an essential ingredient of the statistical theory envisioned in \cite{Monaghan1_1976}.

\subsection*{Acknowledgments}

It is a pleasure to thank L. Lederer, N. Leigh, V. Manwadkar, S. Mazumdar, A. Ori, A. Schiller, U. Smilansky, N. Stone, R. Shir and A. Trani for discussions. I thank Sara Kol for linguistic editing help. 

This work is based in part on ideas on statistical predictions for the double pendulum, a chaotic mechanical system, developed by the author in 2016, initiated by teaching a course on Analytical Mechanics, and tested against simulations in  \cite{KolMarmor2016}.

\newpage
\appendix

\section{Marginalization ignoring absorptivity}
\label{sec:unit_absrp}

In this appendix we present the marginal outcome distributions while ignoring absorptivity, namely assuming $\cE=1$. While in general this a crude approximation, and the distributions are presented only for reference, the escape probability is likely accurate since $\cE$ averages away in this case.

\presub {\bf Marginal distributions.} The distribution of decay rates is given by (\ref{dGm_s}). After integrating over the azimuthal angle with respect to $\vec{L}$ (see figure \ref{fig:E_L_space} (b)) as well as over $\phi_F\, \phi_p$, all three of which are uniformly distributed, we arrive at 
\bea
 d\bGm_s  &=&  \frac{\bsig_\chi}{(2 \pi)^4} \, d\Gamma_s \label{bGm_Gm} \\
 		&=& dl_s(\eps) \; \frac{l_\perp}{l\, l_F} d^2 l\, d^2 l_F\,  \delta^{(2)}\(\vec{l}+\vec{l}_F-\vec{L}\) 
\label{dGm_s2} 
 \eea
 where $d\bGm_s$ is related to $d\Gamma_s$ by a constant which cancels out in probabilities \eqref{dP}; $d^2 l:= dl_\parallel dl_\perp$; $l_\parallel$ is the component of $\vec{l}$ in the direction of $\vec{L}$ and $l_\perp$ is the magnitude of the perpendicular component (see figure \ref{fig:E_L_space} (b)), and similarly for $d^2 l_F$. 

\vspace{0.5cm}


Before we study the general distribution it is instructive to start by studying the limits mentioned in section \ref{sec:spec_pred} where the marginal distributions simplify.

\noindent {\bf $L=0$ case}.  Here the distribution of $\vec{l}$ is isotropic and we have \be
d\bGm_s = 2\, dl_s(\eps) \, dl ~.
\ee
 
Marginalizing through elementary integrations we obtain \bea
	 d\bGm_s  &=& 2\,  l_s(\eps)\, dl_s(\eps)   \non
 	  		&=& 2\, (l_0 - l)\, dl \non			
\bGm_s    &=&  l_0^{~2} 
 \label{L0_marg}
\eea
where the first line gives the binary energy distribution in the range $-\infty \le \eps \le E$ gotten by integration $dl$; the second line is the distribution of binary angular momentum magnitude in the range $0 \le l \le l_s$ where $l_0$ is defined in \eqref{def:l0}; and the third line is the escape probability gotten by double integration. 

The average binary energy for escaper $s$ is determined from the energy distribution through \be
 \left< \eps \right> = \frac{\int \eps\, d\bGm}{\int d\bGm} ~.
 \ee
In the $L=0$ limit is diverges logarithmically \be
\left< \eps \right> \sim \int_{-\infty}^E \frac{d\eps}{\eps} \to -\infty ~.
\ee

In order to determine the distribution over binary eccentricity $e$, we first note that  \be
	e^2 = 1- x^2
\ee
where $x^2 = (-2\, \eps\, l^2)/k_s$ and we have suppressed the dependence of $e,x$ on $s$. This way the definition of $x_s$ \eqref{def:xs} becomes $x_s=x_s(\eps=E,l=L)$. Inserting one into the integration in the form $1= \int dx^2\,  \delta\(x^2-(-2E)l^2/k_s\)$ one finds  \be
  d\bGm_s = l_0^{~2}  \frac{e\, de}{\sqrt{1-e^2}}
 \label{L0_e}
\ee 

We note that in this limit displays planar motion and hence an enhancement of conserved charges. Therefore the direction of the outgoing $\vec{l}$ would not be isotropic as implied by (\ref{dGm_s}) and a planar analysis is in place. Still, the total decay rate and other quantities must be continuous in the $L \to 0$ limit and hence the marginalization above should hold.

\presub {\bf $x_s \ge 1$ case}. This holds for the cases simulated in \cite{MLT}. In this case integration over the polar angle $\theta$ (between $\vec{l}$ and $\vec{L}$ -- see fig. \ref{fig:E_L_space}) gives \be
 d\bGm_s = \frac{2}{L} \, dl_s(\eps) \, l\, dl ~. 
\ee

From this we conclude the following marginal distributions \bea
	 d\bGm_s  &=& \frac{l_s^2(\eps)}{L}\, dl_s(\eps)   \non
 	  		&=& \frac{2}{L} (l_0 -l) \, l\, dl    \non
  &=& \frac{2\, l_0^{~3}}{3\, L}  \, e\, de \non
 \bGm_s    &=&   \frac{l_0^{~3}}{3\, L} 
\label{x1_marg}
\eea

The average binary energy is \be
 \left< \eps \right>  = \frac{\int_{-\infty}^E \eps^{-3/2}\, d\eps} {\int_{-\infty}^E \eps^{-5/2}\, d\eps} = 3\, E ~.
\ee

The $l_F$ distribution can also be determined exactly, even though the integration limits for the angular variables of $\vec{l}_F$ are more involved than for $\vec{l}$. We find \be
dP = \frac{3}{2}\, d\tlf \begin{cases}
\tlf^2 	&	0 \le \tlf \le 1 \\
 (2-\tlf)^2	&	1 \le \tlf \le 2	
\end{cases}
\ee
where \be
\tlf := \frac{l_F-(L-l_0)}{l_0} ~.
\label{def:tlf}
\ee

\presub {\bf General case}. Having gone through the two simpler limits, we are prepared to determine the marginal decay rates in the general case. Integration over $\theta$ uses \be
\int_{-1}^1 \frac{d\cos \theta}{\( l^2 + L^2 - 2 l\, L \cos \theta\)^{1/2} } = \frac{2}{\max \{l,L\}}
\ee 
which we note to be the same integral that computes the electrostatic potential of a charged shell of radius $L$ at radial location $l$.

Applying that to (\ref{dGm_s2}) we find the $\eps,\, l$ distribution \be
 d\bGm_s  = dl_s(\eps) \, \frac{2 l\, dl}{\max \{l,L\}} ~.
\label{gen_eps_l}
\ee
Further integrations lead to the marginal distributions \bea
 d\bGm_s	 &=&  l_s(\eps) \, dl_s(\eps)   \begin{cases} 
 2 - (\eps/\eps_{0s})^{1/2}		&  \eps_{0s} \le \eps \le E \\
 (\eps/\eps_{0s})^{-1/2}				&  \eps \le \min \{ \eps_{0s},\, E \}
 \end{cases} \non 
 	  		&=& 2 \, (l_0 - l)\, \frac{l\, dl}{\max \{l,L\}}    \non
 \bGm_s    &=&  \frac{1}{3}\, l_0^{~2} \begin{cases} 
 3 - 3\, x_s + x_s^2		&  0 \le x_s \le 1 \\
 1/x_s 			&  1 \le x_s
 \end{cases}
\label{gen_marg}
\eea
where \be 
\eps_{0s} := \frac{k_s}{(-2 L^2)} ~.
\ee  
and $l_0$ was defined in \eqref{def:l0}.

We note that for given masses and conserved charges, the lighter the mass, the higher the ejection rate and probability (this appears to hold also for three-body problems outside the realm of physics). This can be seen as follows. First, $\Gamma_s$ is monotonous in $k_s$ since the distribution (\ref{dGm_s2}) is proportional to $\sqrt{k_s}$ while its region (\ref{binary_region}) increases with $k_s$. Equivalently, this is seen by inspecting the expression for $\bGm_s$ in \eqref{gen_marg} within each of the two ranges of $x_s$.  Secondly, for given masses, the lighter the escaper, the larger are both $\alpha$ and $\mu$ of the binary and hence also $k_s$. Combining the two monotonicities implies this paragraph's opening sentence.

For  $0 \le x_s \le 1$ the average binary energy is \be
 \left< \eps \right>  =  3\, E \frac{2 \log x_s^{-1} + x_s}{3-3 x_s +x_s^2}~.
\ee
In the range $0 \le x \le 1 2$, $\langle \eps \rangle/E$ is a monotonous decreasing function, as can be confirmed by plotting it  \cite{Mathematica}. 

The distribution of binary eccentricity is \bea
 d\bGm &=& \frac{dx^2}{6} \begin{cases}
 \frac{3\, l_0^{~2}}{x} - \frac{L^2}{x^3} 	& x_s \le x \le 1 	\\[7pt]
 \frac{2}{L}\, l_0^{~3} 		& 0 \le x \le \min \{ x_s, 1 \}
 \end{cases} = \nonumber  \\[7pt] 
  &=& \frac{e\, de}{3} \begin{cases}
  \frac{1}{\sqrt{1-e^2}} \( 3\, l_0^{~2} - \frac{L^2}{1-e^2} \)	& 0 \le e \le e_s \\[7pt]
  \frac{2}{L} \, l_0^{~3}						& e_s \le e \le 1
  \end{cases}
\label{gen_e}
  \eea
where the first expression is in terms of the $x$ variable, where the expression is simpler, and the second is in terms of $e$ with  $e_s:=\sqrt{1-x_s^2}$. 

All quantities in the general case can be seen to tend to their limits above, and all expressions for $\bGm_s$ can be converted to differential decay rates through \eqref{bGm_Gm}.

\bibliographystyle{unsrt}

\begin{thebibliography}{99}

\bibitem{Principia}
I.~Newton,  ``Philosophi{\ae}  Naturalis Principia
Mathematica'' (1687). 



\bibitem{Euler_1767}
L.~Euler,
``De motu rectilineo trium corporum se mutuo attrahentium,''
Novi commentarii academi{\ae}  scientarum Petropolitan{\ae} 11, pp. 144Ð151 (1767),
 in Oeuvres, Seria Secunda tome XXV Commentationes Astronomic{\ae} (p. 286).

\bibitem{Lagrange_1772}
J.~L.~Lagrange,
``Essai sur le Probl\`eme des Trois Corps,''
Prix de l'Acad\'emie Royale des Sciences de Paris, tome IX (1772),
 in vol. 6 of Oeuvres (p. 292).

\bibitem{Poincare_1892}
H.~Poincar\'e,
``Les m\'ethodes nouvelles de la m\'echanique c\'eleste,''
Gauthier-Villars et fils (1892).

\bibitem{Barrow-Green_1996}
J.~Barrow-Green,
 ``Poincar\'e and the three body problem,''
American Mathematical Society (1996). 

\bibitem{Hand_Finch_1998}
L.~N.~Hand and J.~D.~Finch,
``Analytical Mechanics,''
Cambridge University Press (1998). Chapter 11.

\bibitem{LL_stat_phys}
L.~Landau and E.~M.~Lifshitz,
 ``Statistical Physics,''
 Pergamon Press (1959). 
 
 \bibitem{Gibbs1902}
J.~W.~Gibbs,
 ``Statistical Mechanics,''
 Charles Scribner's Sons (1902).


\bibitem{Agekyan_Anosova_1967}
T.~A.~Agekyan, Z.~P.~Anosova,
 ``A study of the dynamics of triple systems by means of statistical sampling,''
 Astron.\ Zh.\ {\bf 44} 1261 (1967).

\bibitem{Standish_1972}
E.~M.~Standish,
``Dynamical evolution of triple star systems - numerical study,''
Astr.\ Astrophys.\ {\bf 21}, 185 (1972).
%
\bibitem{Saslaw_Valtonen_Aarseth_1974}
W.~Saslaw, M.~J.~Valtonen and S.~J.~Aarseth,
`` Gravitational slingshot and structure of extra-galactic radio-sources,''
Astrophys.\ J.\ {\bf 190}, 253 (1974).


\bibitem{Monaghan1_1976} 
  J.~J.~Monaghan,
  ``Statistical-theory of the disruption of three-body systems - I. Low angular momentum,''
  Mon.\ Not.\ Roy.\ Astron.\ Soc.\ {\bf 176}, 63 (1976).
%

 \bibitem{Monaghan2_1976} 
  J.~J.~Monaghan,
  ``Statistical-theory of the disruption of three-body systems - 2. High angular-momentum,''
  Mon.\ Not.\ Roy.\ Astron.\ Soc.\ {\bf 177}, 583 (1976).
 





\bibitem{Monaghan3_1978} 
  P.~E.~Nash and J.~J.~Monaghan,
  ``Statistical-theory of the disruption of three-body systems - 3. 3-dimensional motion,''
  Mon.\ Not.\ Roy.\ Astron.\ Soc.\ {\bf 184}, 119 (1978).

%


\bibitem{Valtonen_book_2006}
M.~J.~Valtonen and H. Karttunen,
``The three-body problem,''
Cambridge University Press (2006).

\bibitem{Valtonen_etal_book_2016}
 M.~J.~Valtonen, J.~Anosova, K.~Kholshevnikov, A.~Myll\"ari, V.~Orlov and K.~Tanikawa,
``The Three-body Problem from Pythagoras to Hawking,''
Springer (2016).

\bibitem{Musielak_Quarles_rev_2014}
 Z.~E.~Musielak and B.~Quarles,
 ``The three-body problem,''
Rep.\ Prog.\ Phys.\ {\bf 77},  065901 (2014) 
doi: 10.1088/0034-4885/77/6/065901
[arXiv:1508.02312[astro-ph.EP]].

  
\bibitem{Stone_Leigh_2019} 
  N.~C.~Stone and N.~W.~C.~Leigh,
  ``A statistical solution to the chaotic, non-hierarchical three-body problem,''
  Nature {\bf 576}, no. 7787, 406 (2019).
  doi:10.1038/s41586-019-1833-8
 
 
\bibitem{Mathematica}
 Mathematica computing system, Wolfram Research.

\bibitem{LL_mechanics}
L.~Landau and E.~M.~Lifshitz,
 ``Mechanics,''
 Pergamon Press (1960). eq. (49.6) in 2nd edition.

\bibitem{Narnhofer_Thirring_1981}
H.~Narnhofer and W.~Thirring,
``Canonical scattering transformation in classical mechanics,''
Phys.\ Rev.\  A {\bf 23}, 1688 (1987).

\bibitem{Fowler1936}
R.~H.~Fowler,
 ``Statistical mechanics,''
 2nd edition, Cambridge University Press, Cambridge (1936).

\bibitem{Jeans1929}
J.~H.~Jeans,
 ``Astronomy and cosmogony,''
 Dover Publications Inc., New York (1929).

\bibitem{Valtonen_1988}
M.~ J.~ Valtonen,
``The general three-body problem in astrophysics,''
Vistas in Astron.\ {\bf 32}, 23 (1988).
doi: 10.1016/0083-6656(88)90395-9


\bibitem{Gaspard_book_1998}
 P.~Gaspard,
{\it ``Chaos, scattering and statistical mechanics''}
Cambridge University Press (1998). 


\bibitem{Ott_Tel_1993}
 E.~Ott and T.~T\'el
``Chaotic scattering: An introduction,''
 Chaos {\bf 3}, 4 (1993).

\bibitem{Seoane_Sanjuan_2013}
 J.~M.~Seoane and M.~A.~F.~Sanjuan,
 ``New developments in classical chaotic scattering,''
 Rep.\ Prog.\ Phys.\ {\bf 76}  016001 (2013).

\bibitem{Boyd_McMillan_1993}
P.~T.~Boyd and S.~L.~W.~McMillan,
``Chaotic scattering in the gravitational three-body problem,''
Chaos {\bf 3}, 507 (1993). 
https://doi.org/10.1063/1.165956. 

 

\bibitem{Charac_wiki}
Wikipedia, {\it Characteristic function (probability theory)}. 




\bibitem{Valtonen_1974}
M.~ J.~ Valtonen,
``Statistics of three body experiments,''
in ``The Stability of the Solar System and of Small Stellar Systems,'' symposium proceedings
ed. Y. Kozai, p. 211, Reidel, Dordrecht (1974).

\bibitem{Valtonen_Aarseth_1977}
M.~ J.~ Valtonen and S. ~J. ~Aarseth,
``Numerical experiments on the decay of three-body systems,''
Rev. Mex. Astron. Astrofiz. {\bf 3} 163 (1977).

\bibitem{Agekian_Anosova_Orlov_1983}
 T. ~A. ~Agekyan, Zh. ~P. ~Anosova and V. ~V. ~Orlov, 
``Decay time of triple systems,''
Astrophys.\ {\bf 19}, 66 (1983). Translation of Astrofizika {\bf 19}, 111 (1983).
doi: 10.1007/BF01005813


\bibitem{Shevchenko_2010}
 I.~I.~Shevchenko,
``Hamiltonian intermittency and L\'evy flights in the three-body problem,'' 
 Phys.\ Rev.\ E {\bf 81}, 066216 (2010).
  doi:10.1103/PhysRevE.81.066216

\bibitem{Orlov_Rubinov_Shevchenko_2010}
 V.~V.~Orlov, A.~V.~Rubinov and I.~I.~Shevchenko,
``The disruption of three-body gravitational systems: lifetime statistics,'' 
 Mon.\ Not.\ Roy.\ Ast.\ Soc.\ {\bf 408}, 1623 (2010).
 doi = {10.1111/j.1365-2966.2010.17239.x}

 



\bibitem{MLT}
	V.~Manwadkar, A.~A.~Trani and N.~ W.~C.~Leigh,
``Chaos and L\'evy Flights in the Three-Body Problem,''
[arXiv:2004.05475 [astro-ph.EP]].

\bibitem{KolMarmor2016}
B.~Kol, unpublished notes (Jan. 2016).\\
B.~ Kol and A. Marmor,
``Predicting chaos statistically: the double pendulum,''
unpublished paper (Aug. 2016).

\end{thebibliography}

\end{document}